%% file: samplepaper.tex
\documentclass[runningheads]{llncs}
\usepackage{graphicx}
\usepackage{pgfplots}
   \usepackage{booktabs}
   \usepackage{listings}
   \usepackage{xcolor}
   \usepackage{caption}
   \usepackage{subcaption}
   \usepackage{hyperref}
   \usepackage{array}
   \usepackage{tabularx}
    \usepackage{siunitx}
    \usepackage{pgfplots}
    \pgfplotsset{compat=1.18}
    \usepgfplotslibrary{statistics} 
    \usepackage{orcidlink}
\definecolor{codebg}{gray}{0.95}
\definecolor{codeframe}{gray}{0.75}
\definecolor{codekw}{RGB}{0,0,180}
\definecolor{codecomment}{RGB}{60,120,60}
\definecolor{codestring}{RGB}{160,0,0}

\lstdefinestyle{shellstyle}{
  backgroundcolor=\color{codebg},
  frame=single,
  rulecolor=\color{codeframe},
  basicstyle=\ttfamily\footnotesize,
  breaklines=true,
  breakatwhitespace=false,
  numbers=none,
  xleftmargin=6pt,
  xrightmargin=6pt,
  keywordstyle=\color{codekw}\bfseries,
  commentstyle=\color{codecomment}\itshape,
  stringstyle=\color{codestring},
  showstringspaces=false,
  captionpos=b
}

\lstdefinestyle{xmlstyle}{
  backgroundcolor=\color{codebg},
  frame=single,
  rulecolor=\color{codeframe},
  language=XML,
  basicstyle=\ttfamily\footnotesize,
  breaklines=true,
  numbers=none,
  xleftmargin=6pt,
  xrightmargin=6pt,
  keywordstyle=\color{codekw},
  commentstyle=\color{codecomment}\itshape,
  stringstyle=\color{codestring},
  showstringspaces=false,
  captionpos=b,
  morekeywords={cluster,name,code,define,attribute,command,side,type,source,optional}
}

\begin{document}
%
% \title{Contribution Title\thanks{Supported by organization x.}}
% \title{Toward Secure and Interoperable IoMT: Extending the Matter Protocol with a Custom Vital-Signs Cluster}
\title{Demo: Matter-Based IoMT Telemetry and Security Analysis}
%
%\titlerunning{Abbreviated paper title}
% If the paper title is too long for the running head, you can set
% an abbreviated paper title here
%
% \author{First Author\inst{1}\orcidID{0000-1111-2222-3333} \and
% Second Author\inst{2,3}\orcidID{1111-2222-3333-4444} \and
% Third Author\inst{3}\orcidID{2222--3333-4444-5555}}
% %
% \authorrunning{F. Author et al.}
% % First names are abbreviated in the running head.
% % If there are more than two authors, 'et al.' is used.
% %
% \institute{Princeton University, Princeton NJ 08544, USA \and
% Springer Heidelberg, Tiergartenstr. 17, 69121 Heidelberg, Germany
% \email{lncs@springer.com}\\
% \url{http://www.springer.com/gp/computer-science/lncs} \and
% ABC Institute, Rupert-Karls-University Heidelberg, Heidelberg, Germany\\
% \email{\{abc,lncs\}@uni-heidelberg.de}}
%

\author{Muhammad Hassan\inst{1}\orcidID{0000-0002-5713-9658} \and
Masooda Bashir\inst{1} \and
Carl Gunter\inst{1}}

\authorrunning{M. Hassan et al.}

\institute{
University of Illinois Urbana-Champaign, Urbana, IL, USA \\
\email{\{mhassa42, mnb, cgunter\}@illinois.edu}
}

\maketitle              % typeset the header of the contribution
\begin{abstract}
Connected health devices increasingly support health monitoring outside clinical settings, making secure and accessible physiological data important for patients and caregivers. However, device-specific applications and interfaces complicate the integration of readings from different devices. The Matter standard provides standardized device commissioning and protected communication, but its cluster library does not represent the health attribute measurement. We present a preliminary demo study of a Matter-based IoMT testbed with a specific \texttt{HealthVitals} cluster. Implemented in the \texttt{connectedhomeip} SDK on a Raspberry~Pi~5, the cluster exposes simulated heart rate, blood oxygen saturation, and body temperature. We examine commissioning, attribute retrieval, communication under simulated packet loss, and operational traffic through a reference controller. The study demonstrates that this custom telemetry model can operate within the tested Matter configuration without modifying commissioning or secure-session mechanisms. The resulting testbed gives researchers and device developers a concrete platform for investigating health-oriented extensions. It also supports exploration of local telemetry integration that could help patients and caregivers access readings across connected devices.

% Connected health devices and consumer wearables, commonly referred as Internet of Medical Things (IoMT), are now common in everyday care. Many of IoMT devices send data through proprietary systems that lack strong security and do not work well with other devices. This creates a gap between regulated clinical settings and unregulated home use, leaving patients and providers exposed to avoidable privacy and security risks. We address this problem by extending the Connectivity Standards Alliance’s Matter protocol with a custom \texttt{HealthVitals} cluster that encodes heart rate, blood oxygen saturation, and body temperature on a commodity prototype testbed. Our demonstration prototype uses Matter’s SDK tooling to integrate the new cluster without changing standard commissioning, session security, or access control. Evaluation shows that the extended device commissions successfully, exposes IoMT telemetry with correct semantics and stable latency, and preserves encrypted transport and fabric-scoped access enforcement. These results suggest that Matter can serve as a practical foundation for secure, interoperable IoMT devices, benefiting patients, clinicians, and IoT vendors by providing a common, standards-based pathway for health telemetry in cyber-physical systems.

% \keywords{First keyword  \and Second keyword \and Another keyword.}
\keywords{Internet of Medical Things \and Matter protocol \and Cyber-physical systems \and IoT security \and Interoperability \and Security \and Access control}
\end{abstract}
%
%
%-------------------------------------------------------------------------------
\section{Introduction and Research Objective}
\label{sec:introduction}
\input{1_Introduction}
\section{System Design}
\label{sec:sysdesign}
\input{4_SystemDesign}
\section{Workflows and Security Scenarios}
\label{sec:eval}
\input{5_Evaluations}
\section{Discussion and Future Roadmap}
\label{sec:discuss}
\input{6_Discussion}

\section{Conclusion}
\label{sec:conclusion}

This demo illustrates that the CSA’s Matter protocol can be extended to support Internet of Medical Things (IoMT) telemetry without weakening its security or interoperability guarantees. We designed and implemented a custom \texttt{HealthVitals} cluster that encodes heart rate, blood oxygen saturation, and body temperature, integrated it into the open-source Matter SDK, and deployed the resulting firmware on a commodity Raspberry~Pi device. Our evaluation showed that the extended node successfully completes the standard PASE/CASE commissioning sequence, exposes IoMT-relevant attributes with correct semantics and stable latency, and preserves Matter’s encrypted transport and fabric-scoped access-control behavior.

These results provide a concrete feasibility proof that Matter protocol's  mechanism and data-model hierarchy are sufficient to carry health telemetry in a research-grade IoMT setting. By reusing Matter’s commissioning, session-security, and access-control primitives, the prototype offers a standards-based alternative to proprietary stacks that dominate today’s consumer health ecosystem. For patients and clinicians, this points toward IoMT devices that can participate in shared, cryptographically enforced fabrics rather than isolated silos. For device vendors and CPS practitioners, it suggests a practical path to add health-oriented functionality to existing Matter ecosystems with minimal changes to protocol semantics. We view this work as a first step toward a formal family of medical-device clusters within the CSA specification and as a foundation for future studies that combine real biosensors, larger-scale fabrics, and adversarial testing in health-critical cyber-physical systems.

\paragraph{\textbf{Acknowledgement.}}
This work was supported in part by the National Science Foundation (NSF) under grant~1955228.

%
% ---- Bibliography ----
%
% BibTeX users should specify bibliography style 'splncs04'.
% References will then be sorted and formatted in the correct style.
%
\bibliographystyle{splncs04}
\bibliography{mybibliography}
%
% \begin{thebibliography}{8}
% \bibitem{ref_article1}
% Author, F.: Article title. Journal \textbf{2}(5), 99--110 (2016)

% \bibitem{ref_lncs1}
% Author, F., Author, S.: Title of a proceedings paper. In: Editor,
% F., Editor, S. (eds.) CONFERENCE 2016, LNCS, vol. 9999, pp. 1--13.
% Springer, Heidelberg (2016). \doi{10.10007/1234567890}

% \bibitem{ref_book1}
% Author, F., Author, S., Author, T.: Book title. 2nd edn. Publisher,
% Location (1999)

% \bibitem{ref_proc1}
% Author, A.-B.: Contribution title. In: 9th International Proceedings
% on Proceedings, pp. 1--2. Publisher, Location (2010)

% \bibitem{ref_url1}
% LNCS Homepage, \url{http://www.springer.com/lncs}. Last accessed 4
% Oct 2017
% \end{thebibliography}
\end{document}

%% file: 1_Introduction.tex
Connected health devices, also referred to as Internet of Medical Things (IoMT) are increasingly becoming popular and used outside clinical settings. These devices, such as include wearables, pulse oximeters, and connected thermometers, collect health and physiological measurements \cite{el2025internet,nithyavani2025comprehensive}. These devices often use different protocols and standards making interoperability difficult. As a result, a person using different IoMT devices may need separate smartphone applications or controllers to manage and view their health data. This makes it challenging for the user to view and analyze health and wellness data in a unified system \cite{nithyavani2025comprehensive}. Therefore, a smart and connected device protocol that can port the data in a common format so the devices can communicate and be interoperable is needed. Additionally, the protocol should protect the sensitive health data during device setup and routine network exchange \cite{sun2019security}. This can provide a single application and controller for easier interpretation and usability for IoMT devices. 

Matter is an IP-based standard developed for communication among Internet of Things (IoT). It defines a device data model, commissioning procedures, and protected operational sessions \cite{csa_2022,csa_2022_secPr}. These mechanisms make it useful to implement for local health telemetry. However, the Matter cluster library does not define a standard representation for the health and physiological data.

Existing Matter testbeds have examined commissioning and traffic capture across Wi-Fi, Thread, and BLE \cite{mangar2024designing}. Other work has studied deployment on development platforms and provided methods for implementing custom clusters \cite{zegeye2023connected,mangarenabling}. Building on these efforts, we examine a custom cluster in a health-telemetry setting. We focus on how a controller retrieves its attributes after commissioning and how the resulting operational traffic appears in a network capture.

We implement a manufacturer-specific \textit{HealthVitals} cluster using the \textit{connectedhomeip} SDK on a Raspberry~Pi~5. It exposes simulated heart rate, blood oxygen saturation (SpO\textsubscript{2}), and body temperature to a \textit{chip-tool} controller. We report attribute-read behavior, read completion under simulated packet loss, and observations from traffic capture and network-fabric life-cycle tests. Our objective is to assess how this custom telemetry model operates within Matter's existing commissioning and operational workflow.

%% file: 4_SystemDesign.tex
\label{sec:system-design}

\begin{figure}[htbp]
  \centering
  \includegraphics[
    width=0.65\columnwidth,
    alt={A Raspberry Pi 5 runs the HealthVitals cluster on Endpoint 1.
    A Linux host runs chip-tool. Both hosts communicate over local Wi-Fi,
    using PASE during commissioning and CASE for operational reads.}
  ]{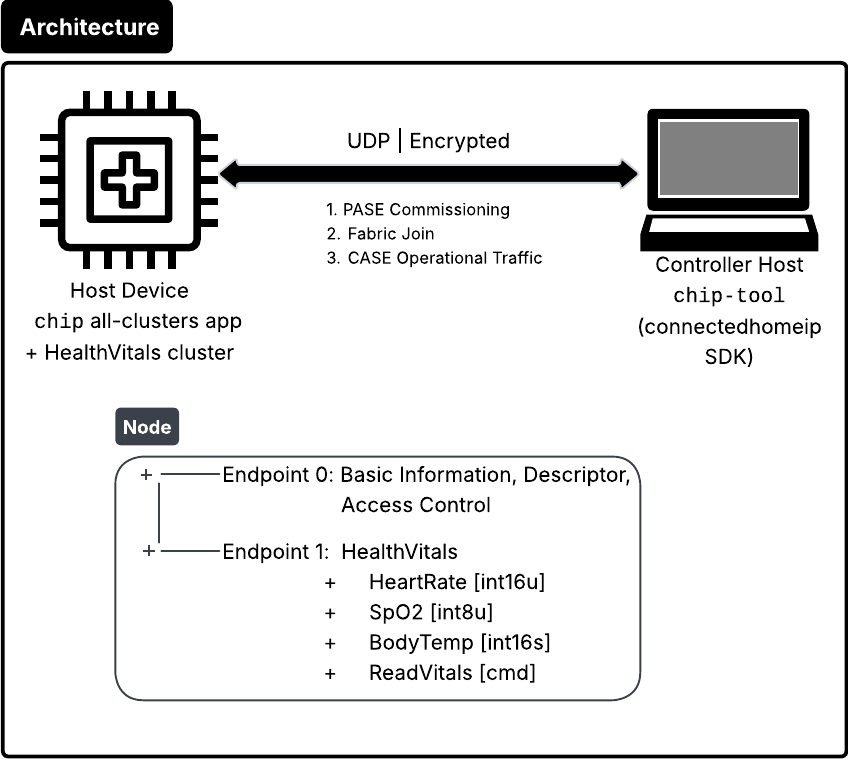}
  \caption{Architecture of the Matter-based IoMT testbed.}
  \label{fig:testbed-arch}
\end{figure}

The testbed is build using a Raspberry~Pi~5 endpoint and a Linux controller host connected over a local Wi-Fi network. The endpoint runs a modified \textit{chip-all-clusters-app} from the \textit{connectedhomeip} SDK. The controller runs \textit{chip-tool}, built from the same SDK source tree. The hosts exchange telemetry locally, without a cloud relay or Thread border router. Simulated readings replace physical biosensors, allowing application behavior to be tested independently of sensor acquisition. Figure~\ref{fig:testbed-arch} shows the architecture and the placement of the custom cluster.

The Matter standard organizes device functionality into endpoints containing clusters, which define attributes, commands, and events \cite{csa_2022}. We added a manufacturer-specific \textit{HealthVitals} cluster to Endpoint~1, alongside the application's existing Matter functionality. The cluster uses identifier \textit{0xFFF1FC01} and exposes three readable attributes with explicit type and unit conventions. \textit{HeartRate} is an unsigned 16-bit integer representing beats per minute, with simulated values between 40 and 200. \textit{SpO2} is an unsigned 8-bit integer representing oxygen saturation, with simulated values between 85\% and 100\%. \textit{BodyTemp} is a signed 16-bit integer representing temperature in tenths of a degree Celsius. Its stored values range from 360 to 380, corresponding to 36.0--38.0$^\circ$C.

The application initializes the attributes using bounded random variation and regenerates their values when the controller invokes \textit{ReadVitals}. The command refreshes the stored state; subsequent attribute reads retrieve the resulting values through Matter's interaction model. The current schema represents three simulated values rather than timestamped measurement records. This design separates telemetry generation from communication, providing a straightforward integration point for future sensor acquisition and measurement metadata. 
We declared the cluster schema in the SDK's data-model files and regenerated its associated artifacts using ZAP. We then implemented initialization and command-handling routines and compiled the endpoint application using the GN/Ninja build system. The controller accesses the cluster through identifier-based commands, using the cluster, attribute, and endpoint identifiers to specify each request. Device onboarding uses Passcode-Authenticated Session Establishment (PASE), while operational reads use Certificate Authenticated Session Establishment (CASE) \cite{csa_2022_secPr}. The extension therefore adds application-level telemetry while retaining the SDK's existing commissioning and secure-session implementation.

%% file: 5_Evaluations.tex
\label{sec:workflows}

We commissioned the endpoint using \texttt{chip-tool}'s \texttt{already-discovered} mode, supplying its network address and port directly. This procedure bypassed mDNS discovery during initial pairing. Commissioning established a PASE session, provisioned operational credentials, and added the endpoint to the controller's fabric. The controller subsequently established CASE-secured communication for operational attribute reads \cite{csa_2022_secPr}. Each \texttt{read-by-id} request specified the cluster, attribute, node, and endpoint identifiers. The \texttt{ReadVitals} command regenerated the simulated values before subsequent reads retrieved the updated state.

We assessed attribute-read behavior across two runs, checking response types and values against the implemented schema. The first run contained ten sequential \texttt{HeartRate} reads; the second contained five reads of each attribute. All 25 requests returned correctly typed values within the configured ranges. For the second run, mean read times after excluding the first trial were 293.0~ms, 292.8~ms, and 293.0~ms. These values correspond to \texttt{HeartRate}, \texttt{SpO2}, and \texttt{BodyTemp}, respectively. The first trial was excluded consistently across attributes; its additional delay was not attributed to a specific protocol mechanism.

We examined request completion under network disruption using \texttt{tc netem} with configured packet-loss rates of 0\%, 5\%, and 15\%. Ten attribute reads were issued at each rate, and all 30 requests completed successfully. This experiment measured request completion rather than latency under loss or the number of retransmissions. It therefore provides an initial observation of application behavior under the selected impairment conditions.

We captured traffic on the endpoint's WLAN interface using \texttt{tcpdump} while the commissioned controller issued attribute reads. The captured operational exchanges did not expose the simulated vital-sign values in plaintext. This observation is consistent with Matter's protection of operational messages through secure sessions \cite{csa_2022_secPr}. The analysis concerned network traffic generated by these reads, rather than data stored or displayed on either host.

We also examined operational discovery under two controller credential conditions, contrasting them with the commissioned controller's successful reads. A fresh \texttt{chip-tool} instance with a separate credential store timed out during mDNS resolution after 45.145~s. Following \texttt{RemoveFabric}, a read using the removed fabric's controller state similarly timed out after 45.109~s. Both attempts stopped at operational discovery, before CASE establishment or an application-layer attribute request. These observations describe discovery behavior in the tested configurations, rather than an authenticated request denied by an access-control list. They provide a starting point for separately examining discovery, session establishment, and authorization in future experiments.

%% file: 6_Discussion.tex
\label{sec:discussion}

The prototype provides an implementation path for representing health measurement values within Matter's existing data model and operational workflow. Its custom cluster adds health-related attributes while retaining the SDK's commissioning and secure-session implementation. Successful reads and the operational traffic capture provide a basis for further study of local health telemetry. The future extension plans to build on this implementation through richer measurement records, targeted security experiments, and independent controller support.

The \texttt{HealthVitals} schema can be extended with measurement timestamps, validity states, and sensor provenance. These fields would help controllers distinguish recent measurements from stale readings and identify their source. Connecting physical sensors would also enable evaluation of sampling intervals, acquisition delays, and updates to the stored attributes. Together, these extensions would support a measurement pipeline linking sensor acquisition to controller retrieval.

The security evaluation can be extended by testing authorization within an established CASE session. An authenticated controller with restricted permissions would issue reads against \texttt{HealthVitals}, while endpoint logs record the interaction-layer response. This would separate authorization behavior from operational discovery and session establishment. Additional scenarios could examine permission changes, existing sessions after fabric removal, and deliberately misconfigured access-control policies. These experiments would assess how Matter's existing controls govern broadly access to the custom telemetry \cite{csa_2022_secPr}.

An independently implemented controller would provide a further test of the cluster's encoding, units, and command behavior. Larger fabrics and longer experiments could examine concurrent reads, network impairment, and performance across multiple endpoints. Finally, mappings to IEEE~11073 and HL7~FHIR could connect the cluster's local representation to established health-data models \cite{ieee_11073,fhir}. These steps would provide evidence for assessing a reusable Matter health-telemetry profile beyond the current testbed.